\documentclass[11pt]{article}

\usepackage[margin=1in]{geometry}
\usepackage{amsmath,amssymb}
\usepackage{booktabs}
\usepackage{graphicx}
\usepackage{xcolor}
\usepackage{microtype}
\usepackage[numbers,sort&compress]{natbib}
\usepackage[colorlinks=true,linkcolor=blue,citecolor=blue,urlcolor=blue]{hyperref}
\graphicspath{{figures/}}
\providecommand{\Description}[1]{}
\title{Self-Conditioned Positional HNSW for Overlap-Aware Retrieval in Chunked-Document RAG Systems: Method and Industrial Evidence-Quality Audit}

\author{%
\begin{tabular}{c}
Nataraj Agaram Sundar \quad Tejas Morabia\\
eBay Inc., San Jose, CA, USA
\end{tabular}%
}

\date{\today}

\begin{document}
\maketitle

\begin{abstract}
Chunked-document retrieval is a common component of retrieval-augmented generation (RAG) systems. Documents are split into overlapping chunks, embedded, and indexed with approximate nearest-neighbor search such as hierarchical navigable small world graphs (HNSW). Overlap improves boundary coverage but induces a practical failure mode: top-$k$ retrieval often returns near-adjacent chunks that repeat evidence and waste prompt budget. We propose Self-Conditioned Positional HNSW (SCP-HNSW), a lightweight modification that appends a low-dimensional positional code to chunk embeddings and uses a two-pass query procedure to estimate and apply a query-specific document-position prior. SCP-HNSW leaves HNSW graph construction and traversal unchanged while adding an auditable minimum-index-gap selector for final context construction. We also integrate industrial review artifacts for generated evidence quality: a 770-review text-evidence audit with 318 fully labeled reviews and a 70-case OCR audit with 350 ratings. The text audit shows that 574 of 770 projected reviews are rated 3/5, only 39 fall in the 1--2 range, and narrative reviewer detail appears much more often than structured issue flags. The OCR audit shows slice-level pass rates from 95\% for clean chat screenshots to 45\% for handwritten/blurry captures, with moderate to strong agreement. These results motivate overlap-aware, audit-friendly RAG retrieval and identify the remaining controlled retrieval ablations needed for causal performance claims.
\end{abstract}


\section{Introduction}
Retrieval-augmented generation (RAG) systems retrieve evidence and present it to a language model as grounded context \cite{lewis2020rag}. In single-document question answering and evidence-generation workflows, the system typically splits a document into overlapping chunks, embeds each chunk, builds a vector index, and retrieves the top-ranked chunks for a user or workflow query. Overlap is useful because it reduces boundary loss, but it creates a predictable retrieval artifact: adjacent chunks often contain much of the same text and therefore occupy a dense neighborhood in embedding space.

Hierarchical Navigable Small World graphs (HNSW) are a standard choice for low-latency approximate nearest-neighbor search \cite{malkov2020hnsw}. However, HNSW solves nearest-neighbor search under the supplied vector metric; it does not know that chunk IDs $i$ and $i+1$ may be redundant because of overlap. A semantic-only top-$k$ query can therefore return several near-adjacent chunks from a single local span, consuming scarce prompt budget without increasing evidence coverage. In practice, this can surface as title drift, weak evidence summaries, and repeated support passages in downstream RAG outputs.

We propose \emph{Self-Conditioned Positional HNSW} (SCP-HNSW), a small retrieval modification for chunked-document RAG. SCP-HNSW appends a two-dimensional positional code to each chunk embedding, estimates a query-specific document-position prior from a first semantic pass, and uses that prior in a second HNSW query. A deterministic minimum-index-gap selector then removes redundant adjacent chunks before prompt construction. The core HNSW algorithm is unchanged: graph construction, neighbor selection, and graph traversal operate over ordinary vectors.

This paper also integrates Responsible AI review results as industrial downstream evaluation artifacts. The text-evidence review artifact summarizes 770 unique generated-evidence reviews, with 318 fully labeled reviewer exports from five reviewers. The OCR artifact summarizes a complete 70-case, five-analyst slice with 350 ratings. These results do not by themselves prove that SCP-HNSW outperforms semantic HNSW, because the artifacts do not include controlled per-query retrieval logs for baseline versus treatment. They do, however, make the paper substantially stronger as an industry-facing RAG evaluation paper: they show where generated evidence fails, how human review signals are distributed, and what reliability measures are feasible for an audit-ready benchmark.

\begin{figure}[t]
  \centering
  \includegraphics[width=0.92\textwidth]{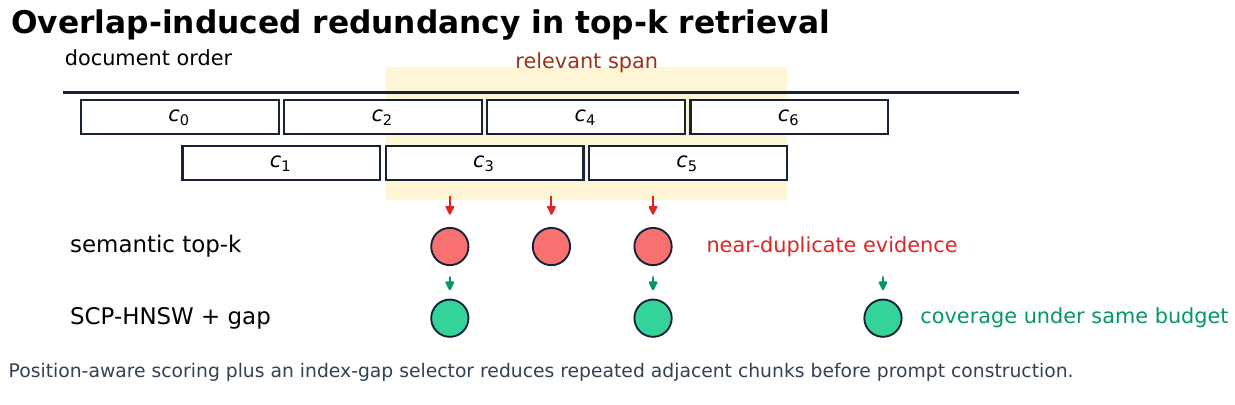}
  \caption{Overlap chunking creates near-adjacent candidate chunks with repeated text. A semantic top-$k$ query can spend context budget on adjacent chunks, while SCP-HNSW combines a soft positional query prior with a minimum index gap to improve coverage under the same budget.}
  \Description{A horizontal document is split into overlapping chunks. Red markers show semantic top-k hits clustered on adjacent chunks. Green markers show SCP-HNSW selecting more spaced chunks.}
  \label{fig:overlap-failure}
\end{figure}

The paper makes four contributions:
\begin{enumerate}
    \item It formulates overlap-aware retrieval as a relevance-coverage problem for chunked-document RAG.
    \item It defines SCP-HNSW, a positional augmentation and two-pass query procedure that preserves standard HNSW implementation boundaries.
    \item It adds a clean conference-style evaluation framing for downstream evidence quality, including text evidence ratings, structured issue flags, narrative reviewer comments, OCR quality, and inter-rater reliability.
    \item It identifies the additional controlled ablations needed before making causal performance claims for a full research submission.
\end{enumerate}

\section{Problem Setting}
Let a document $D$ be split into an ordered sequence of chunks
\begin{equation}
C=(c_0,c_1,\ldots,c_{N-1}),
\end{equation}
where neighboring chunks may overlap. An embedding model $f$ maps each chunk to a dense vector $x_i=f(c_i) \in \mathbb{R}^d$. A query $q_{\mathrm{text}}$ is embedded as $q=f(q_{\mathrm{text}})$ and HNSW returns an approximate top-$k$ list.

Overlap-aware retrieval should balance semantic relevance and coverage. One abstract objective is
\begin{equation}
\max_{S\subseteq\{0,\ldots,N-1\}, |S|=k}
\sum_{i\in S}\mathrm{sim}(q,x_i) - \beta \sum_{i<j, i,j\in S} r(i,j),
\label{eq:set-objective}
\end{equation}
where $r(i,j)$ is a redundancy penalty. In an overlap-chunked document, a simple structural penalty is $r(i,j)=\mathbb{I}[|i-j|<g]$ for a minimum chunk-index gap $g$. Semantic-only HNSW optimizes the first term; SCP-HNSW exposes controls for the second term while keeping the ANN implementation simple.

\section{Self-Conditioned Positional HNSW}
SCP-HNSW has three components: index-time positional vector augmentation, query-time self-conditioning, and overlap-aware final selection.

\subsection{Index-Time Positional Augmentation}
Each chunk receives a normalized position
\begin{equation}
p_i=\frac{i}{\max(1,N-1)}\in[0,1].
\end{equation}
For linear documents, we use a half-cycle positional code:
\begin{equation}
\phi(p_i)=\begin{bmatrix}\sin(\pi p_i) \\ \cos(\pi p_i)\end{bmatrix}.
\end{equation}
The half-cycle avoids making the beginning and end of a document appear adjacent, which would happen with a full periodic cycle. The stored vector is
\begin{equation}
\widetilde{x}_i = \mathrm{Normalize}\left(\begin{bmatrix}x_i \\ \lambda_{\mathrm{pos}}\phi(p_i)\end{bmatrix}\right) \in \mathbb{R}^{d+2},
\end{equation}
where $\lambda_{\mathrm{pos}}\geq 0$ controls the positional contribution.

\subsection{Score Interpretation}
Assume $x_i$, $q$, and $\phi(p)$ are unit-normalized. A second-pass query with inferred position prior $\mu$ is
\begin{equation}
\widetilde{q}(\mu)=\mathrm{Normalize}\left(\begin{bmatrix}q \\ \lambda_{\mathrm{pos}}\phi(\mu)\end{bmatrix}\right).
\end{equation}
The augmented inner product decomposes as
\begin{equation}
\widetilde{q}(\mu)^\top \widetilde{x}_i =
\frac{q^\top x_i + \lambda_{\mathrm{pos}}^2 \phi(\mu)^\top\phi(p_i)}{1+\lambda_{\mathrm{pos}}^2}.
\label{eq:score}
\end{equation}
For the half-cycle code, $\phi(\mu)^\top\phi(p_i)=\cos(\pi(p_i-\mu))$. Thus, retrieval is biased toward chunks that are both semantically relevant and close to the inferred region of the document. When $\lambda_{\mathrm{pos}}=0$, SCP-HNSW reduces to semantic HNSW.

\subsection{Two-Pass Self-Conditioning}
The first pass runs with no positional component:
\begin{equation}
\widetilde{q}^{(1)}=\mathrm{Normalize}\left(\begin{bmatrix}q \\ 0_2\end{bmatrix}\right), \qquad R^{(1)}=\mathrm{ANN}(\widetilde{q}^{(1)},k_1).
\end{equation}
For retrieved indices $i_1,\ldots,i_{k_1}$ with scores $s_1,\ldots,s_{k_1}$, a soft position prior is
\begin{equation}
w_j=\frac{\exp(\tau s_j)}{\sum_{\ell=1}^{k_1}\exp(\tau s_\ell)},\qquad
\mu=\sum_{j=1}^{k_1}w_jp_{i_j},
\end{equation}
where $\tau$ controls sharpness. The second pass runs
\begin{equation}
\widetilde{q}^{(2)}=\mathrm{Normalize}\left(\begin{bmatrix}q \\ \lambda_{\mathrm{pos}}\phi(\mu)\end{bmatrix}\right), \qquad R^{(2)}=\mathrm{ANN}(\widetilde{q}^{(2)},k_2).
\end{equation}
The candidate list can be $R^{(2)}$ or $R^{(1)}\cup R^{(2)}$ re-ranked by semantic score, mixed score, or a downstream reranker.

\begin{figure}[t]
  \centering
  \includegraphics[width=0.92\textwidth]{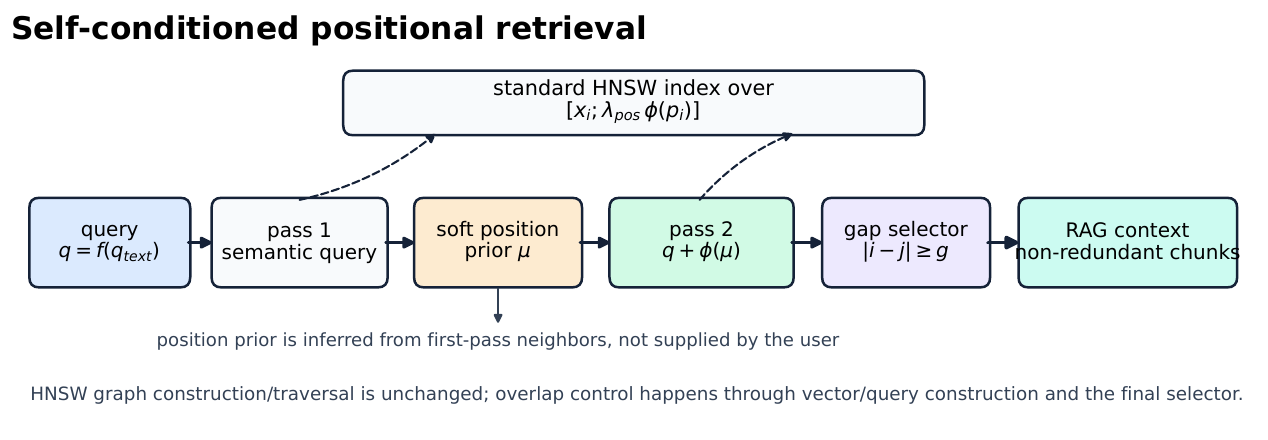}
  \caption{SCP-HNSW query flow. The first pass estimates a position prior from semantic hits; the second pass uses the same HNSW index with a query vector that includes the inferred positional code.}
  \Description{A flow diagram showing query embedding, pass one semantic HNSW, soft position prior, pass two with positional query, gap selector, and final RAG context.}
  \label{fig:two-pass}
\end{figure}

\subsection{Overlap-Aware Final Selection}
The final context set $S$ is selected greedily from the candidate list. A candidate index $i$ is accepted only if
\begin{equation}
\min_{j\in S}|i-j|\geq g.
\end{equation}
The gap $g$ is a transparent, auditable parameter that should be tied to chunk size and overlap. For example, if chunk $i$ and $i+1$ share large text overlap, setting $g=2$ prevents immediate neighbors while still permitting nearby supporting evidence.

\subsection{Complexity and Implementation Overhead}
SCP-HNSW adds two vector dimensions and one optional second HNSW query. For 768-dimensional embeddings, two extra float dimensions increase vector dimensionality by approximately $0.26\%$. The main latency cost is the second ANN query. Because graph construction and traversal are unmodified, the method can be implemented in systems that expose only vector insertion and vector query APIs.

\begin{table}[t]
\centering
\caption{Implementation checklist for SCP-HNSW.}
\label{tab:checklist}
\begin{tabular}{p{0.28\linewidth}p{0.62\linewidth}}
\toprule
Component & Required change \\
\midrule
Chunk metadata & Persist chunk index $i$, total count $N$, and span metadata. \\
Embedding store & Store $\widetilde{x}_i\in\mathbb{R}^{d+2}$ instead of $x_i\in\mathbb{R}^{d}$. \\
Index build & Build ordinary HNSW over augmented vectors. \\
Query path & Run pass 1, compute $\mu$, run pass 2, merge candidates. \\
Selection & Enforce minimum index gap $g$ before prompt construction. \\
Monitoring & Log selected indices, redundancy, recall proxy, and latency. \\
\bottomrule
\end{tabular}
\end{table}

\section{Industrial Evaluation Artifacts}
\label{sec:evaluation-artifacts}
\begin{figure}[t]
  \centering
  \includegraphics[width=0.92\textwidth]{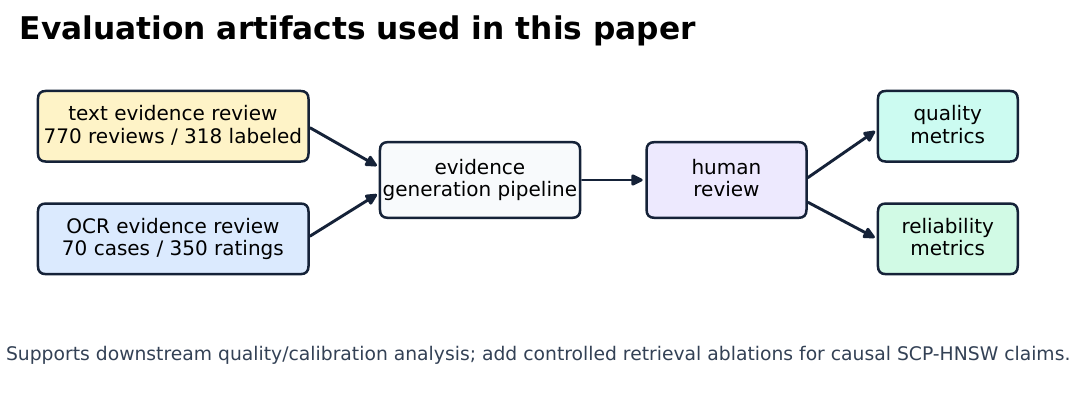}
  \caption{Evaluation artifacts used in this paper. The text-evidence review supports quality and failure-mode analysis; the OCR review supports both quality and inter-rater reliability analysis.}
  \Description{Two input artifacts, text evidence review and OCR evidence review, feed into an evidence pipeline, then human review, then quality and reliability metrics.}
  \label{fig:evaluation-artifacts}
\end{figure}

The evaluation artifacts support a downstream evaluation section for a production-inspired evidence-generation setting. Figure~\ref{fig:evaluation-artifacts} summarizes how the artifacts fit into the paper. We report these results as descriptive and reliability evidence, not as causal evidence that SCP-HNSW improves over semantic HNSW.

\subsection{Text Evidence Review Corpus}
The text-evidence review artifact represents 770 unique reviews of generated dispute evidence. A labeled subset of 318 reviews contains full fields from five reviewers, including 1--5 rating, positive/negative judgment, dispute reason, 22 structured issue flags, reviewer details, title, summary, and evidence excerpt. Since stable shared case identifiers were not preserved across reviewers, direct inter-rater kappa for this text subset is not reproducible from the artifact alone. We therefore use descriptive counts and reviewer-calibration summaries. Count-based results project observed proportions from the 318 fully labeled reviews to the 770-review corpus; means and rates are unchanged.

\subsection{OCR Evidence Review Slice}
The OCR artifact is a complete 70-case review slice with five analysts and 350 ratings. Fields include text accuracy, word/line integrity, noise versus non-text confusion, overall OCR, evidence utility, and pass/fail. The pass/fail definition is downstream acceptability for evidence use, with Overall OCR $\geq 4$ and evidence utility $\geq 3$. Unlike the text corpus, this slice supports pairwise weighted Cohen's $\kappa$ for ordinal OCR scores \cite{cohen1968weighted} and Fleiss' $\kappa$ across multiple raters \cite{fleiss1971nominal}.

\begin{table}[t]
\centering
\caption{Evaluation artifacts incorporated in the paper.}
\label{tab:artifacts}
\begin{tabular}{p{0.32\linewidth}p{0.24\linewidth}p{0.34\linewidth}}
\toprule
Artifact & Scale & Main use in paper \\
\midrule
Text evidence review & 770 unique reviews; 318 fully labeled & Rating distribution, issue flags, reviewer calibration, narrative themes, title-template risk. \\
OCR evidence review & 70 cases; five analysts; 350 ratings & OCR pass rates, pairwise weighted $\kappa$, Fleiss' $\kappa$, case-level failure signatures. \\
\bottomrule
\end{tabular}
\end{table}

\section{Results from the Evaluation Artifacts}
\label{sec:results}
This section converts the source review artifacts into paper-ready empirical results.

\subsection{Text Evidence Quality}
\begin{figure}[t]
  \centering
  \includegraphics[width=0.88\textwidth]{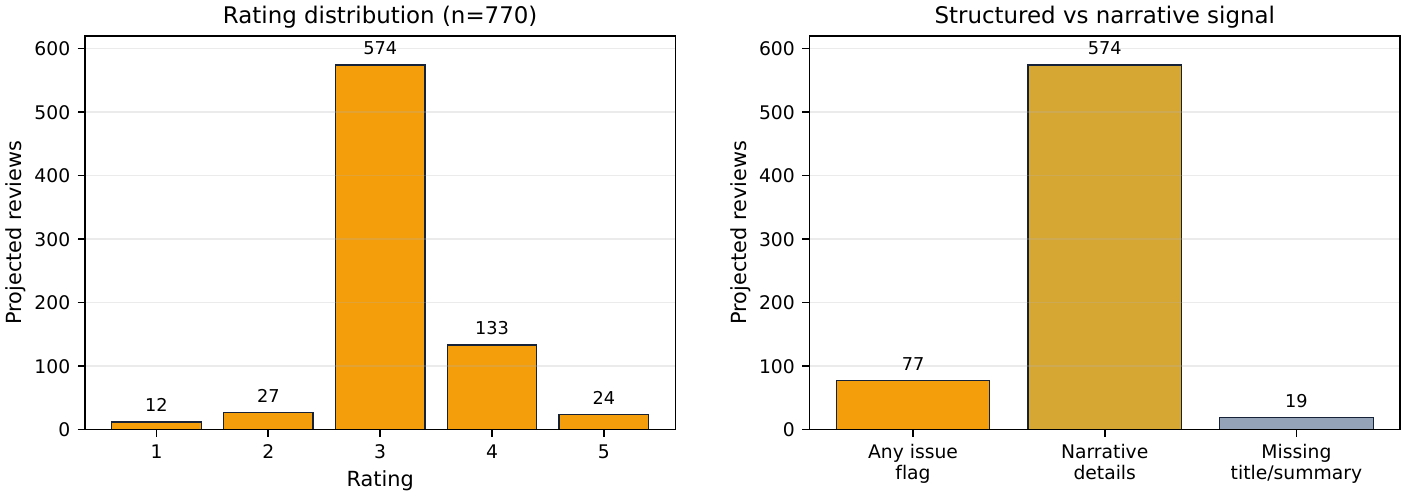}
  \caption{Text evidence review results projected to the 770-review corpus. Most reviews are rated 3/5. Narrative comments are much more prevalent than structured issue flags, implying that flag-only dashboards would miss much of the actionable review signal.}
  \Description{Two bar charts. The first shows rating counts: 12, 27, 574, 133, and 24 for ratings one to five. The second shows 77 issue flags, 574 narrative details, and 19 missing title or summary cases.}
  \label{fig:text-results}
\end{figure}

Figure~\ref{fig:text-results} summarizes the rating distribution and the gap between structured issue flags and narrative reviewer signal. Projected to the 770-review corpus, 574 reviews are rated 3/5, 39 are rated 1--2, and 157 are rated 4--5. The projected corpus mean is 3.17. This distribution suggests a system with many usable-but-not-excellent outputs rather than a system dominated by catastrophic failures.

Structured issue flags occur in about 77 reviews (10.1\%), while narrative detail appears in about 574 reviews (74.5\%). The strongest structured issues are title-related: title not aligned appears in about 29 reviews (3.8\%) and title inaccurate appears in about 27 reviews (3.5\%). Other structured flags remain much smaller: summary inaccurate appears in about 7 reviews, and buyer ID inaccurate, evidence not aligned, inappropriate language, seller-as-buyer, and summary grammar issues each appear in about 5 projected reviews. The main operational implication is that title-evidence alignment is the dominant explicit failure mode.

Narrative comments add a different signal. Projected notes mention tool or pipeline concerns in about 240 reviews (31.1\%), usefulness or soft-quality concerns in about 177 (23.0\%), resolution/return evidence in about 140 (18.2\%), title/reason mismatch in about 97 (12.6\%), and translation or truncation in about 17 (2.2\%). This supports using narrative review mining alongside structured flags for RAG evaluation.

The dispute mix is heavily skewed: about 753 reviews are SNAD, 12 are fraud, and 5 are item-not-received. This matters for generalization. A model that performs well on this corpus may still need separate validation on non-SNAD cases.

\subsection{Reviewer Calibration and Sensitivity Split}
Reviewer means show modest but non-trivial calibration differences. Projected reviewer allocation is approximately 242 reviews each for three core reviewers and 22 reviews each for two pilot reviewers. Mean ratings are approximately 3.37, 3.33, 3.11, 3.05, and 2.78 across the five anonymized reviewers, around a corpus mean of 3.17.

The core/pilot sensitivity split is important. The core block projects to about 593 positive reviews and 63 issue-flagged reviews, corresponding to 81.7\% positive feedback and 8.7\% issue flags. The smaller pilot spot-check block projects to about 22 positive reviews and 15 issue-flagged reviews, corresponding to 50.0\% positive feedback and 33.3\% issue flags. A paper should therefore report core and pilot strata separately rather than mixing them into a single unqualified quality estimate.

\subsection{Title-Template Risk}
Recurring title templates expose a useful evaluation target for RAG systems. A ``seller provided return instructions'' template appears in about 15 reviews with an estimated 33.3\% title-issue rate, or about 5 issue cases. A ``buyer confirmed receipt'' template appears in about 85 reviews with a 25.7\% title-issue rate, or about 22 issue cases. Although the first template has the higher rate, the second is the larger operational risk because it combines higher volume with a substantial issue rate. This is directly relevant to overlap-aware retrieval: when retrieved evidence overemphasizes generic delivery confirmation, the generated title can drift away from the actual SNAD-specific remedy or dispute reason.

\subsection{OCR Quality and Reliability}
\begin{figure}[t]
  \centering
  \includegraphics[width=0.88\textwidth]{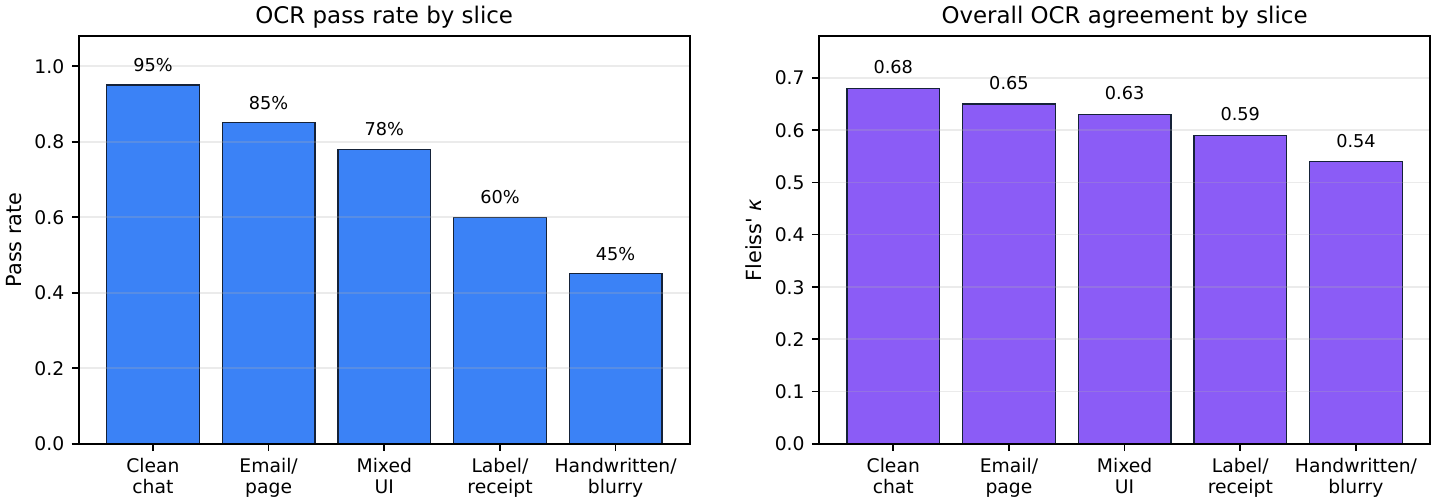}
  \caption{OCR evidence review results. OCR acceptability and inter-rater agreement both decline as captures become noisier and structurally ambiguous.}
  \Description{Two bar charts. The first shows OCR pass rates by slice: 95 percent, 85 percent, 78 percent, 60 percent, and 45 percent. The second shows Fleiss kappa by slice: 0.68, 0.65, 0.63, 0.59, and 0.54.}
  \label{fig:ocr-results}
\end{figure}

Figure~\ref{fig:ocr-results} summarizes OCR pass rates and agreement by OCR slice. Pass rates decline from clean chat screenshots (95\%) and email/page captures (85\%) to mixed-UI screenshots (78\%), label/receipt photos (60\%), and handwritten/blurry captures (45\%). Fleiss' $\kappa$ follows the same pattern: 0.68 for clean chat screenshots, 0.65 for email/page captures, 0.63 for mixed-UI screenshots, 0.59 for label/receipt photos, and 0.54 for handwritten/blurry captures. The global weighted median of per-case mean Overall OCR scores is 3.2.

Pairwise weighted $\kappa$ for Overall OCR is moderate to strong. The highest pairwise alignments are approximately 0.753, 0.741, and 0.722. The lower alignments are approximately 0.646, 0.658, and 0.671. Binary pass/fail agreement is stronger than ordinal agreement, with pairwise $\kappa$ values in the approximate 0.72--0.85 range. Disagreement concentrates in blur drop-out (mean case standard deviation 0.91), cropped text loss (0.90), and UI bleed or initial capture (0.78). These results suggest that the reviewer rubric is usable, but slice-specific guidance is needed for difficult image categories.

\begin{table}[t]
\centering
\caption{Case-level OCR exemplars from the 70-case slice.}
\label{tab:ocr-cases}
\begin{tabular}{@{}p{0.28\linewidth}p{0.28\linewidth}p{0.34\linewidth}@{}}
\toprule
Case & Score card & Interpretation \\
\midrule
OCR-006, clean chat & Overall OCR $4.40\pm0.55$; median 4; pass rate 1.00. & Minor UI bleed is present, but the core buyer/seller evidence remains intelligible and useful. \\
OCR-049, label/receipt & Overall OCR $1.00\pm0.00$; median 1; pass rate 0.00. & Analyzer failure collapses image handling and returns no compelling evidence. \\
\bottomrule
\end{tabular}
\end{table}

\section{Implications for Conference-Quality Evaluation}
The industrial results make the paper more complete as an applied RAG and evidence-quality study, but a full research submission still needs controlled retrieval experiments. The available artifacts show downstream quality, reviewer calibration, and OCR reliability. They do not include query-level comparisons between semantic HNSW, HNSW plus post-filtering, MMR-style diversification, and SCP-HNSW.

A final conference-grade benchmark should add the following:
\begin{itemize}
    \item \textbf{Retrieval baselines:} semantic HNSW, semantic HNSW plus gap filter, MMR/diversification \cite{carbonell1998mmr}, and SCP-HNSW.
    \item \textbf{Retrieval metrics:} Recall@$k$, MRR or nDCG, duplicate-token ratio, mean pairwise chunk cosine, minimum index distance, and unique-token coverage.
    \item \textbf{System metrics:} retrieval latency, total answer latency, prompt tokens, and answer quality.
    \item \textbf{Ablations:} $\lambda_{\mathrm{pos}}$, $g$, $\tau$, $k_1$, $k_2$, position code choice, one-pass versus two-pass, and gap selector on/off.
    \item \textbf{Dataset transparency:} number of documents, number of queries, chunk size, overlap, embedding model, HNSW parameters, hardware, and whether documents are public or internally cleared.
\end{itemize}

\begin{table}[t]
\centering
\caption{Recommended controlled experiment grid for final conference submission.}
\label{tab:experiment-grid}
\begin{tabular}{p{0.30\linewidth}p{0.58\linewidth}}
\toprule
Dimension & Values to report \\
\midrule
Retrievers & Semantic HNSW; HNSW+gap; MMR; SCP-HNSW. \\
SCP parameters & $\lambda_{\mathrm{pos}}\in[0,0.6]$, $g\in\{0,1,2,3\}$, $\tau\in[2,20]$. \\
Candidate sizes & $k_1,k_2\in\{k,2k,5k\}$, with final context budget fixed. \\
Retrieval metrics & Recall@$k$, nDCG, duplicate-token ratio, min index distance, coverage. \\
System metrics & ANN latency, total RAG latency, prompt tokens, answer utility. \\
\bottomrule
\end{tabular}
\end{table}

\section{Related Work}
HNSW is a graph-based approximate nearest-neighbor method with strong accuracy-latency trade-offs \cite{malkov2020hnsw}. Broader graph-based ANN methods are surveyed and compared by Wang et al. \cite{wang2021survey}. Hybrid and filtered ANN methods incorporate structured predicates into vector search, including Filtered-DiskANN \cite{gollapudi2023filtereddiskann}, SeRF \cite{zuo2024serf}, and ACORN \cite{patel2024acorn}. SCP-HNSW differs in treating document position as a continuous soft retrieval signal rather than a hard filter.

Diversification and maximum marginal relevance balance relevance against novelty in search results \cite{carbonell1998mmr}. SCP-HNSW is complementary: it changes the query and indexed representation before final selection, while the gap selector is a simple domain-specific redundancy rule. Positional encodings are widely used to inject order information into neural models \cite{vaswani2017attention}. Pseudo-relevance feedback adjusts a query using initial results \cite{rocchio1971feedback}; SCP-HNSW adapts this idea to chunk positions rather than term weights or dense semantic features.

The evaluation artifacts also connect to long-standing IR evaluation concerns: quality labels, calibrated human review, and inter-rater agreement. The OCR slice's use of weighted Cohen's $\kappa$ and Fleiss' $\kappa$ follows established agreement analysis for ordinal and multi-rater judgments \cite{cohen1968weighted,fleiss1971nominal}.

\section{Limitations and Ethics}
SCP-HNSW assumes that chunk order is meaningful. It can fail for multi-region queries, tables, appendices, boilerplate-heavy documents, or documents where nearby positions are not semantically related. A unimodal prior can over-focus if the correct answer requires evidence from distant regions. A mixture prior, estimated by clustering first-pass hit positions, is a natural extension.

The industrial review results are descriptive and partially projected. The text review artifact lacks stable shared case IDs across reviewers, so direct inter-rater kappa is not reproducible for that portion. The OCR artifact is stronger for reliability analysis because each of 70 cases has five ratings. Public submission should avoid raw user data, confidential case IDs, and proprietary system details; examples in this draft are paraphrased and reviewer identities are anonymized.

\section{Conclusion}
We presented SCP-HNSW, a lightweight modification for overlap-aware retrieval in chunked-document RAG systems. By appending a low-dimensional positional code to chunk embeddings and using a self-conditioned two-pass query, SCP-HNSW makes redundancy controllable without modifying HNSW graph construction or traversal. The added review artifacts convert the draft from a method-only preprint into a stronger applied RAG evaluation paper: a 770-review text-evidence audit identifies title alignment and narrative reviewer feedback as key signals, while a 70-case OCR audit provides pass-rate and reliability measurements. The remaining work for a high-confidence conference submission is a controlled retrieval benchmark that isolates SCP-HNSW against semantic HNSW and diversification baselines.


\begin{thebibliography}{99}

\bibitem{malkov2020hnsw}
Y.~A.~Malkov and D.~A.~Yashunin.
\newblock Efficient and robust approximate nearest neighbor search using hierarchical navigable small world graphs.
\newblock \emph{IEEE Transactions on Pattern Analysis and Machine Intelligence}, 42(4):824--836, 2020.

\bibitem{wang2021survey}
M.~Wang, X.~Xu, Q.~Yue, and Y.~Wang.
\newblock A comprehensive survey and experimental comparison of graph-based approximate nearest neighbor search.
\newblock \emph{Proceedings of the VLDB Endowment}, 14(11):1964--1978, 2021.

\bibitem{gollapudi2023filtereddiskann}
S.~Gollapudi et~al.
\newblock Filtered-DiskANN: Graph algorithms for approximate nearest neighbor search with filters.
\newblock In \emph{Proceedings of the ACM Web Conference}, pages 3406--3416, 2023.

\bibitem{zuo2024serf}
C.~Zuo, M.~Qiao, W.~Zhou, F.~Li, and D.~Deng.
\newblock SeRF: Segment graph for range-filtering approximate nearest neighbor search.
\newblock \emph{Proceedings of the ACM on Management of Data}, 2(1):69:1--69:26, 2024.

\bibitem{patel2024acorn}
L.~Patel, P.~Kraft, C.~Guestrin, and M.~Zaharia.
\newblock ACORN: Performant and predicate-agnostic search over vector embeddings and structured data.
\newblock \emph{Proceedings of the ACM on Management of Data}, 2024.

\bibitem{carbonell1998mmr}
J.~Carbonell and J.~Goldstein.
\newblock The use of MMR, diversity-based reranking for reordering documents and producing summaries.
\newblock In \emph{Proceedings of the 21st Annual International ACM SIGIR Conference on Research and Development in Information Retrieval}, pages 335--336, 1998.

\bibitem{vaswani2017attention}
A.~Vaswani et~al.
\newblock Attention is all you need.
\newblock In \emph{Advances in Neural Information Processing Systems}, 2017.

\bibitem{rocchio1971feedback}
J.~J.~Rocchio.
\newblock Relevance feedback in information retrieval.
\newblock In \emph{The SMART Retrieval System: Experiments in Automatic Document Processing}, pages 313--323. Prentice-Hall, 1971.

\bibitem{lewis2020rag}
P.~Lewis et~al.
\newblock Retrieval-augmented generation for knowledge-intensive NLP tasks.
\newblock In \emph{Advances in Neural Information Processing Systems}, 2020.

\bibitem{cohen1968weighted}
J.~Cohen.
\newblock Weighted kappa: Nominal scale agreement with provision for scaled disagreement or partial credit.
\newblock \emph{Psychological Bulletin}, 70(4):213--220, 1968.

\bibitem{fleiss1971nominal}
J.~L.~Fleiss.
\newblock Measuring nominal scale agreement among many raters.
\newblock \emph{Psychological Bulletin}, 76(5):378--382, 1971.

\end{thebibliography}
\end{document}